\documentclass[useAMS,usenatbib]{mn2e}
\usepackage{url,times,graphicx,amsmath,amsfonts,amssymb,aas_macros,color,epsfig,varioref,subfigure,comment,natbib}
\usepackage[normalem]{ulem}
\usepackage{float} %required for the placement specifier H -> not working!
\usepackage{ae,aecompl}

\usepackage{tikz}
\newcommand{\Tab}[1]{Table~\ref{#1}}

\newcommand{\Fig}[1]{Fig.~\ref{#1}}
\newcommand{\hMpc}{{\ifmmode{h^{-1}{\rm Mpc}}\else{$h^{-1}$Mpc}\fi}}
\newcommand{\hGpc}{{\ifmmode{h^{-1}{\rm Mpc}}\else{$h^{-1}$Gpc}\fi}}
\newcommand{\hkpc}{{\ifmmode{h^{-1}{\rm kpc}}\else{$h^{-1}$kpc}\fi}}

\newcommand{\hMsun}{{\ifmmode{h^{-1}{\rm {M_{\odot}}}}\else{$h^{-1}{\rm{M_{\odot}}}$}\fi}}
\newcommand{\hmsun}{{\ifmmode{h^{-1}{\rm {M_{\odot}}}}\else{$h^{-1}{\rm{M_{\odot}}}$}\fi}}
\newcommand{\Msun}{{\ifmmode{{\rm {M_{\odot}}}}\else{${\rm{M_{\odot}}}$}\fi}}
\newcommand{\kpc}{{\ifmmode{{\rm kpc}}\else{kpc}\fi}}
\def\hMpc{$h^{-1}\,{\rm Mpc}$}
\def\hkpc{$h^{-1}\,{\rm kpc}$}
\def\kms{{ \rm km} $s^{-1}$}
\def\LCDM{\ensuremath{\Lambda}CDM}

\newcommand{\mlg}{{\ifmmode{M_{\rm tot}}\else{$M_{\rm tot}$}\fi}}
\newcommand{\vtan}{{\ifmmode{v_{\rm tan}}\else{$v_{\rm tan}$}\fi}}
\newcommand{\vrad}{{\ifmmode{v_{\rm rad}}\else{$v_{\rm rad}$}\fi}}

\usepackage{comment}
\usepackage{soul,xcolor}

\def\mmw{$M_{\rm MW}$}
\def\mm31{$M_{\rm M31}$}
\def\mmto{$M_{\rm M31}$}

\title
[Estimation of the masses in the Local Group]
{Estimation of the masses in the Local Group by Gradient Boosted Decision Trees}
\author[Carlesi Edoardo]
{
	Edoardo Carlesi $^1$ \thanks{E-mail: ecarlesi83@gmail.com}, 
	Yehuda Hoffman $^1$  \thanks{E-mail: hoffman@huji.ac.il},
	Noam I Libeskind $^{2,3}$\\
$^1$Racah Institute of Physics, Hebrew University, Jerusalem, Israel\\
$^2$Leibniz Instit\"ut f\"ur Astrophysik Potsdam (AIP), An der Sternwarte, Potsdam, Germany\\
$^{3}$University of Lyon, UCB Lyon 1, CNRS/IN2P3, IUF, IP2I Lyon, France\\
\setlength{\topmargin}{-1.5cm}
}

\begin{document}

\date{Submitted XXXX XXX XXXX}

\pagerange{\pageref{firstpage}--\pageref{lastpage}} \pubyear{2021}

\maketitle

\label{firstpage}

%\setstcolor{blue}
\setstcolor{red}
%\st{Text}

%%%%%%%%%%%%%%%%%%%%%%%%%%%%%%%%%%%%%%%%%%%%%%%%%%%

\begin{abstract}

Our goal is to estimate the  mass of the Local Group (LG) and the individual masses of its  primary galaxies,the  M31 and the Milky Way (MW).
We do this by means of a supervised machine learning algorithm, the gradient boosted decision trees (GBDT) and using the observed distance and relative velocity of the two as input parameters.
 The GBDT is applied to a sample of  2148 mock LGs drawn from a set of 5 dark matter (DM)-only simulations, ran withing the standard \LCDM\ cosmological model.  The selection of the mock LGs  is guided by a LG model, which defines such objects.  The role of the observational uncertainties of the input parameters is gauged by applying the model to an ensemble of mock LGs pairs whose observables are these input parameters perturbed by their corresponding observational errors. Finally the observational data of the actual LG is used to infer its relevant masses. Our main results are the sum and the individual    masses of the MW and M31: $M_{tot} = 3.31 ^{+0.79}_{-0.67} $, $M_{MW}=1.15^{+0.25}_{-0.22}$ and $M_{M31}=2.01^{+0.65}_{-0.39} \ \ \times 10^{12}M_{\odot}$  (corresponding to the median and the 1st and 3rd quartiles). The ratio of the masses is $M_{M31}/M_{MW}=1.75^{+0.54}_{-0.28}$, where by convention the M31 is defined here to be the more massive of the two halos. 
\end{abstract}

%%%%%%%%%%%%%%%%%%%%%%%%%%%%%%%%%%%%%%%%%%%%%%%%%%%%%%%%%%%%%%%%%%%%%%%%%%%%%%%%%%


%%%%%%%%%%%%%%%%%%%%%%%%%%%%%%%%%%%%%%%%%%%%%%%%%%%%%%%%%%%%%%%%%%%%%%%%%%%%%%%%%%

\section{Introduction}\label{sec:intro}

In recent years, a substantial development in cosmology has been driven by the increase in quality and quantity of observational
data in the nearby Universe. In fact, thanks to the sharp improvement in cosmological simulations, it has been possible to test to 
higher degrees of accuracy, the predictions of the standard $\Lambda$ Cold Dark Matter (\LCDM) model on small scales. 
In particular, observations of the Local Group (LG) of galaxies has shaped galaxy formation theories and their links to cosmology 
at large \citep{Boylan-Kolchin:2012, Zavala:2012, Tollerud:2014, Elahi:2015, Penzo:2016, Garaldi:2016, Carlesi:2017b}. 
However, the  estimation  of the  masses of the LG as well as its two main constituents  - the Milky Way (MW) and Andromeda (M31) galaxies - 
is still an open an unresolved issue.

The so-called \emph{timing argument} model \citep{Kahn:1959, Lynden-Bell:1981},
even with the simplifying assumptions of a purely radial orbits and a first time infall for the MW and M31,
provides a good proxy to the dynamics of the LG. It applies the Newtonian two-body problem model to integrate the equations of motion of the LG from the Big Bang to the present epoch. This approach can be extended by including the Large Magellanic Clouds \citep{Penarrubia:2016}
and Dark Energy \citep{Partridge:2013}. Other methods of mass estimation usually rely on cosmological simulations,  e.g. in \citet{Carlesi:2016b}
where it was highlighted the effect of the uncertainty in the transverse velocity of the MW-M31 system, in \citet{McLeod:2017} who estimated the impact
of the large scale structure on the mass value and the likelihood-free inference method of \citet{Lemos:2021}, who used \emph{Gaia} data \citep{Gaia:2016} 
in combination with large dark matter simulations.

The   \emph{timing argument} model assumes a very simplified presentation of the LG - two point-like objects of time invariant masses and unperturbed by the rest of the universe. The mass of the LG is  defined as the sum of the masses of the two bodies. The actual LG is much more complicated than that. It is made of two main extended bodies of time evolving masses as well as dozens of smaller satellite galaxies; and its dynamics is affected by the tidal field exerted by the external universe.  The notion of the `mass of the LG' is not well defined and it strongly depends on   how the LG is defined. We try here to stay close to the spirit of the \emph{timing argument} model and use the sum of the masses of the   MW and M31 as a proxy to the mass of the LG.

Machine learning (ML) algorithms have been shown to be  extremely useful tools in cosmology and astrophysics.
Given their capability of modelling non-linear relationships, they have been used to address a vast array of different issues, such as galaxy classification \citep[][]{Lahav:1995}, 
cosmological model selection \citep[]{Merten:2019, Peel:2019}, dark matter halo properties \citep[]{Lucie-Smith:2018, Lucie-Smith:2019, Lucie-Smith:2020}, 
local group mass \citep[]{McLeod:2017, Lemos:2021}, the relation between gas and dark matter distribution \citep[]{Machado:2020}, photometric redshifts \citep[]{Collister:2004} and galaxy properties in 
surveys \citep[]{Mucesh:2021}, among others.
In the present context ML   refers to a class of computational techniques that assign   output values to a set of input parameters, called in the ML jargon \emph{features}. The unique character of ML algorithms is that they bypass the standard approach of solving analytically or more often numerically a set of non-linear equations. Instead statistical hidden relations are used to predict a desired solution. The term \emph{supervised  machine learning} refers to a wide class of ML algorithms that  are trained and validated by using  large datasets  for which the desired output values are known - either by detailed calculations or by extra observations or measurements - for the input data.

We aim here at estimating the masses of the LG and the MW and M31 galaxies from the observed kinematics of the LG within the framework of supervised ML, applying the \emph{gradient boosted decision trees} (GBDT) algorithm to the observed values of the distance between the MW and the M31 galaxies and their relative radial and tangential velocities to calculate the above masses. The supervised nature of the ML requires the construction of an ensemble of mock LGs, that closely emulate the actual observed LG, for which the full dynamics of such objects is available, in particular both the relative distances  and velocities of the mock MW and M31 and their masses are know known,
A sample of mock LGs is constructed here from a set of cosmological  DM-only $N$-body simulations.
DM halos are used as proxies for the MW and M31 galaxies and the   mass of the LG is  approximated here  by \mlg = \mmw\ + \mmto, while the individual masses are the corresponding halo masses. The training and validation of the GBDT to the ensemble of mock LGs yields the desired non-linear mapping from the input features to the desired masses. 
The simulations are conducted within the standard \LCDM\ cosmological model and the mock LGs are defined by an assumed LG model (see below)
We are interested here in the posterior  distribution of the output masses  given the observed input features and the prior distribution of their associated uncertainties and within the framework of the \LCDM\ and the LG models. This is done by constructing an ensemble of Monte Carlo realizations of possible  observed features and using the GBDT algorithm to map it into an ensemble of possible outcomes. This is used to estimate  the mean and scatter of the predicted masses.

The paper is structured as follows. \S \ref{sec:LGmodel} presents the LG model. The construction of the ensemble of mock LGs, that are used as the training and validations sets, is summarized in \S \ref{sec:mock_LGs}. Only a   brief introduction to the vast issue of supervised ML, in general, and the GBDT methodology is given here (\S \ref{sec:SML}). The application  of the GBDT algorithm to the  actual LG is presented in \S \ref{sec:actual_LG}.  A summary and general discussion conclude the paper (\S \ref{sec:disc}).

\section{The Local Group model}
\label{sec:LGmodel}

A selection of mock LGs from simulations needs to be preceded by a discussion of what defines it in simulations. Namely, one needs to outline the of  rules that capture the main relevant properties of  the observed LG and use  it  for  selecting mock LGs from  the simulations. 
This is the  so-called Local Group Model (see \citealt{Carlesi:2016b}).
In the context  of the dynamical modeling of the LG the model focuses and the following properties of the two main DM halos that constitute  the  LG:  masses, mass ratio, isolation, separation and relative    velocity.
In this case, the parameters and their allowed intervals  are: 1. Halo masses: ($M_{200}$, that is the total mass enclosed within the radius at which the density is 200 times the critical one) within $(0.5 - 4.0)\times10^{12}$\hMsun; 2. Mass ratio: \mmto\ : \mmw\ $<4$; 3.  Separation: $r = (0.4-1.3)$\hMpc; 4.   Isolation, i.e. no other halo of mass $\geq$ \mmw\ within 2\hMpc; 5.  Radial velocity: \vrad $<0$ \kms
A further assumption is made that the M31 galaxy is the more massive member of the LG. This choice does not affect the outcome of the analysis and is made for the sake of mathematical convenience  and as a way of labelling the most massive halo of the LG. Yet, it   is motivated by  \cite{Karachentsev:2009} and \cite{Diaz:2014}, though a more massive  Milky Way  cannot be excluded \citep{Gottesman:2002, Ibata:2004}.

The results presented here  are valid under the assumption of \LCDM\ and the LG models. In a Bayesian language the these  models serve as the prior which conditions the results that follow. 

A general comment concerning the $h^{-1}$ scaling is due here, where $h=H_0/100\ km\ s^{-1} Mpc^{-1}$ and $H_0$ is Hubble's constant. As far as data drawn from simulations is concerned, such as masses and distances, the $h^{-1}$ is used. Data from observations of the actual  LG appears without that scaling.

\section{The training and validation datasets}
\label{sec:mock_LGs}

Supervised ML algorithms are based on the availability of a given set of data for which the sought after results are known. In the astrophysical context where the LG is a unique object one needs to  resort to selecting mock LGs from simulations, for which all the information - including the masses of the MW- and M31-like halos - is known. The training and validation datasets are  taken here from  
 a set of five DM only $N$-body simulations (introduced in \citealt{Carlesi:2019}) 
with $1024^3$ particles within a 100 \hMpc\ box with Planck-I cosmological parameters \citep{Planck:2013}.
The halo catalogs were obtained with \texttt{AHF} \citep{Knollmann:2009}, which uses an adaptive mesh refinement technique
to identify halo centers as density peaks in the matter field and iteratively removes gravitationally unbound particles. 
The mass of a halo extracted from the numerical simulations  is defined by $M_{200}$, namely the mass inside a sphere within which the mean density is 200 times the critical density, an attribute of the AHF halo finder \citep[for details]{Knollmann:2009}.

The LG model provides a very generous definition of the LG that allows to build up a robust sample of 2148 objects to be used for the training and validation of the ML algorithm.
As mentioned above, the LG model does not distinguish between the MW and M31 halos and therefore we somewhat arbitrarily follow the commonly held convention that  \mmw $<$ \mmto.  The median   and the 1st and 3rd  quartiles of  the distribution of the mock LGs are:  
\mmw$=0.64_{-0.17}^{+0.41}\times 10^{12}$\hMsun\ and \mmto$1.39_{-0.54}^{+0.94}\times 10^{12}$\hMsun.
The  mock LGs are designed to emulate the   dynamics of the observed LG. As such the data that defines a mock LG consists of the distance ($r$) between the two main halos of the object and their relative radial and tangential velocities  (\vrad\ and  \vtan). These are defined with respect to the center of mass of the halos.

\section{Supervised machine learning   }
\label{sec:SML}

\subsection{Basics}
\label{subsec:basics
}
The problem addressed here is   of an estimation of  \mlg, \mmw, \mmto and \mmw/\mmto given three input parameters that define a LG by means of supervised ML. Our approach adopted  is to calculate the  four output  parameters (results) independently of one another. Namely four independent ML functionals are constructed  as follows:
\begin{eqnarray}
 M_{MW}  & = &  D_{MW }(r, v_{rad},v_{tan})  \nonumber\\
 M_{M31} & = &  D_{M31}(r, v_{rad},v_{tan})  \nonumber\\
 M_{tot} & = &  D_{tot}(r, v_{rad},v_{tan})  \\
r_M\equiv {M_{MW}\over M_{M31}} & = & D_{r_M}(r, v_{rad},v_{tan})   \nonumber
\label{eq:ML-funct}
\end{eqnarray}
The four operators $D_y$ (where $y$ stands for  the individual masses, total mass and for  the mass ratio) are multi-parameters non-linear operators that map from the three input parameters to one output parameter. The structure of these functionals is determined by the rules of the specific assumed ML algorithm and its many free parameters  are determined by non-linear regression  performed on a training dataset  and   tested against  a validation dataset. This is the essence of supervised ML algorithm.  The non-linear regression is performed independently for each one of the functionals of Eqs. 1.

The  supervised ML algorithm used here is the  GBDT  that operates on a combined set of decision trees models. A brief description of   decision trees models in general  and of the particular GBDT method follows.

\subsection{Decision Trees}
\label{subsec:dec-tree}

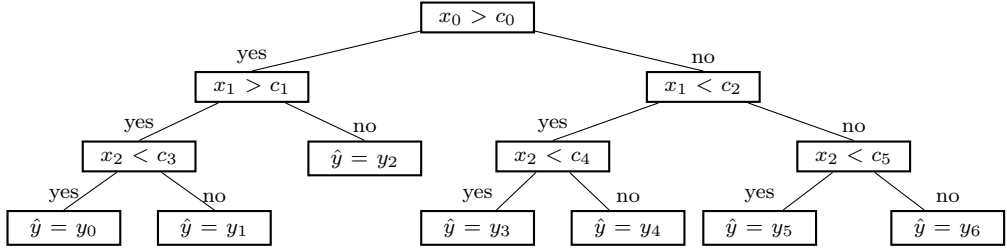
\begin{figure*}
\begin{center}:
    \begin{tikzpicture}[sibling distance=6cm,
      	fact/.style={rectangle, text width=1.3cm, fill=white, text centered, draw=black,thick, anchor=north, text=black},
    	leaf/.style={circle, text width=1.5cm, fill=white, draw=black,thick, text centered, anchor=north, text=black},
    level distance=0.5cm, growth parent anchor=south]
    \node [fact] {$x_0>c_0$}
    child { [sibling distance=3cm] node [fact, label=yes] { $x_1 > c_1$} 
    child { [sibling distance=2cm] node [fact, label=yes]{ $x_2 < c_3$} 
      child { [sibling distance=1cm] node [fact, label=yes] {$\hat{y}=y_0$} }
      child { [sibling distance=1cm] node [fact, label=no] {$\hat{y}=y_1$} }
    }
    child { [sibling distance=3cm] node [fact, label=no]{ $\hat{y}=y_2$} }
    }
    child { [sibling distance=4cm] node [fact, label=no] { $x_1<c_2$}
    child { [sibling distance=2cm] node [fact, label=yes]{ $x_2<c_4$} 
      child { node [fact, label=yes] {$\hat{y}=y_3$} }
      child { node [fact, label=no] {$\hat{y}=y_4$} }
    }
    child { [sibling distance=2.5cm] node [fact, label=no]{ $x_2<c_5$} 
      child { node [fact, label=yes] {$\hat{y}=y_5$} }
      child { node [fact, label=no] {$\hat{y}=y_6$} }
    }
    };
\end{tikzpicture}
  \caption{A graphic presentation of a schematic example of a decision tree defined by  $N=3$ input features, $x_0, ..., x_{N-1}$, $I=6$ nodes and $N_{leaf}=7$ leaves. The $x_n$ is the n-th component of a  given data point,   $c_i $ is the i-th constatnt that defines  the threshold value of the i-th node and $y_{nl}$ is the $nl$-th leave, namely a possible outcome, that the decision tree can predict. 
	}
	\label{fig:tree}

\end{center}
\end{figure*}

A decision tree $d(\textbf{x})$ is a supervised learning algorithm that predicts a  continuous or discrete single target variable $y$ 
from an array of $N$ input features $\textbf{x} = (x_0, ..., x_{N-1})$ which describes each data point. In the particular case considered here we have $N=3$. The input features are processed through a set of nodes - Boolean conditions  with a 'yes' or 'no' outcome - and ends up with one possible outcome out of predetermined possible outcomes. In the language of decision trees methods the possible outcomes are denoted as leaves.
In the case where the physical nature of the outcome $y$ is that of a continuous variable the number of leaves dictates the numerical resolution of the tree. For a discrete target variable $y$, e.g. in a classification problem, the number of leaves equals the number of possible classes. A given decision tree is defined by 
the number and arrangements of the nodes and  leaves    - these are considered the hyper-parameters of the particular tree. A given node consists of the Boolean condition of whether $x_n > c_i$, where $n=0,..., N-1$ and $i=0, ..., I-1$ with $I$ being a hyper-parameter that determines the number of thresholds $c_i$. The decision tree  shown in \Fig{fig:tree}  
is designed for $N=3$ features, $I=6$ nodes and $N_{leaf}=7$ leaves. Such a decision tree can assign one out of 8 possible pre-determined numerical values for the questioned asked, namely $\hat{y}$ accepts one of the possible values of $y_l$ ($l=0, ..., N_{leaf}-1$). Here, an estimate of the total mass of the LG, say, is     $\hat{y}=d_{LG}(r, v_{rad},v_{tan})$. 

The optimization of a given tree proceeds as follows. The hyper-parameters within which the optimization is done are the maximum depth of the tree ($n_{depth}$) and the maximum number of leaves ($N_{leaf}$), which are selected ab initio. The structure of the nodes, ensemble of parameters $\left\{ c_i\right\}$, their number $I$ and the actual number of leaves $N_{leaf}$ are fixed by the optimization, which   is done by the minimization the loss function $L$ defined as the 
mean `distance'  between the actual and the predicted value taken over all members of the training set, 
\begin{equation}
L = {1\over N_{train}}\left( \sum_0^{N_{train}}\left(y_\alpha - \hat{y}_\alpha \right)^2\right)^{1/2},
\end{equation}
where $N_{train}$ is the number of data points in the training set and the sum extends over all all members of that set.

\subsection{Gradient boosted decision trees}

Decision trees are often considered to be `weak learners' - their   simple structure tends to keep their predicted output `close' to the data of the training set, hence they are prone  to over-fitting. 
Further boosting is needed to overcome it. 
The gradient boosted decision trees algorithm \citep[]{Freund:1997, Friedman:2001} is an ensemble method    where several decision trees are combined together to provide a more reliable estimate of the target
output.

In the the  random forest algorithm    an ensemble of  $N_{tree}$ decision trees, all having the same hyper-parameters, is constructed.
The number of the trees and their hyper-parameters are fixed at the onset of the calculation.   The random forest estimator is the ensemble averaged value of the output of the individual trees    defined by:
\begin{equation}
D(x) = \frac{1}{N_{tree}} \sum _{j=1}^{N_{tree}} d_j(x).
\label{eq:D(x)}
\end{equation}

An alternative approach is to start with  one tree  $d_1(x)$ and gradually add successively more trees,  $d_2(x), d_3(x)..., d_{i}(x)$, again all  constructed with the same hyper-parameters.
Given the first $i$ decision trees,  a new estimator $D_i(x)$ is constructed by 
\begin{equation}
    D_j(x) = \frac{1}{j} \sum_{j^\prime=1}^{j} w_{j^\prime} d_{j^\prime}(x),
\label{eq:D(x)}
\end{equation}
whose output is $\hat y_j = D_j(x)$. The weights $w_{j^\prime}$ and the parameters of $d_{j^\prime}(x)$ are adjusted in order to ensure that $L_{j} < L_{j-1}$. 
The procedure is iterated until the desired number of total trees $N_{tree}$ is reached. The hyper-parameters and the number of trees are gauged by the convergence of $L_j$, calculated for the training set.
This is the gradient boosted decision trees algorithm, and it is the method chosen here to estimate the masses of the LG.

The essence of the GBDT algorithm is the non-linear mapping from a set of input features to an outcome, namely, a prediction, whose possible numerical values is spanned by the set of leaves. The structure of a single decision tree or the full GBDT mapping is done by a complex non-linear regression with respect to a loss function in a way that does not easily yield itself to a simple analytical understanding. Yet these algorithms can rank the input features by their importance, namely by the sensitivity of the outcome, i.e. result, to the different features. The output of a single decision tree or the GBDT algorithms is a rearrangement of the  input features by a decreasing order of importance. The output is most sensitive to the first feature and the least to the last one. In a way this is akin to the order of modes in the principal component analysis \citep[PCA;][]{2002nrca.book.....P}. A detailed description on how the importance ranking is done is beyond the scope of the paper. Much like in the PCA case the importance ranking provides insight to the nature of the underlying physical processes, where a low feature importance  means that the desired outcome depends weakly on it. It provides also a tool for the reduction of the dimensionality of the data, and thereby increase the computational efficiency of a given case.

For the specific GBDT algorithm used here the following hyper-parameters need to be specified ab initio:  the number of trees ($N_{tree}$), the maximum depth of the tree ($N_{depth}$) and the minimum number of elements per leaf node $N_{min\_leaf}$. The later parameter refers here to the minimal number of mock LGs, drawn from the training set, so as to avoid over=fitting the algorithm to the training data. Avoiding  such a constraint might lead to the number of leaf nodes being equal to the number of input mock LGs. 
The analysis presented here  has been  using the \texttt{Python} code \texttt{MLEKO}\footnote{MLEKO (Meaning milk in Serbian language) 
stands for Machine Learning Environment for KOsmology and is available at http://github.com/EdoardoCarlesi/MLEKO}, which relies on the 
\texttt{scikit-learn} \citep{scikit-learn} implementation of the GBDT algorithm.

\subsection{Training and testing}
\label{sec:model}

\begin{figure*}
	\begin{center}
		\includegraphics[width=0.4\textwidth]{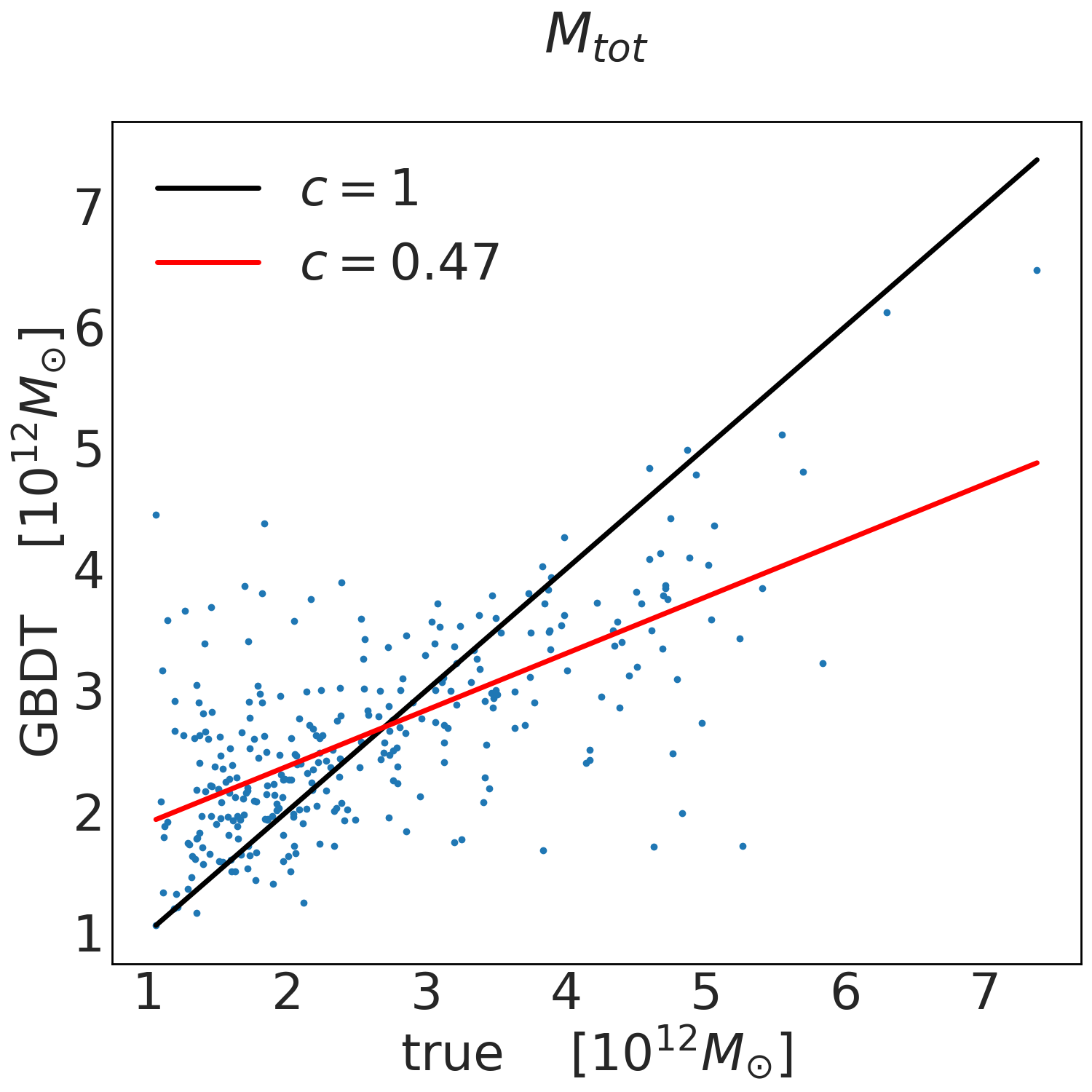}  
		\includegraphics[width=0.4\textwidth]{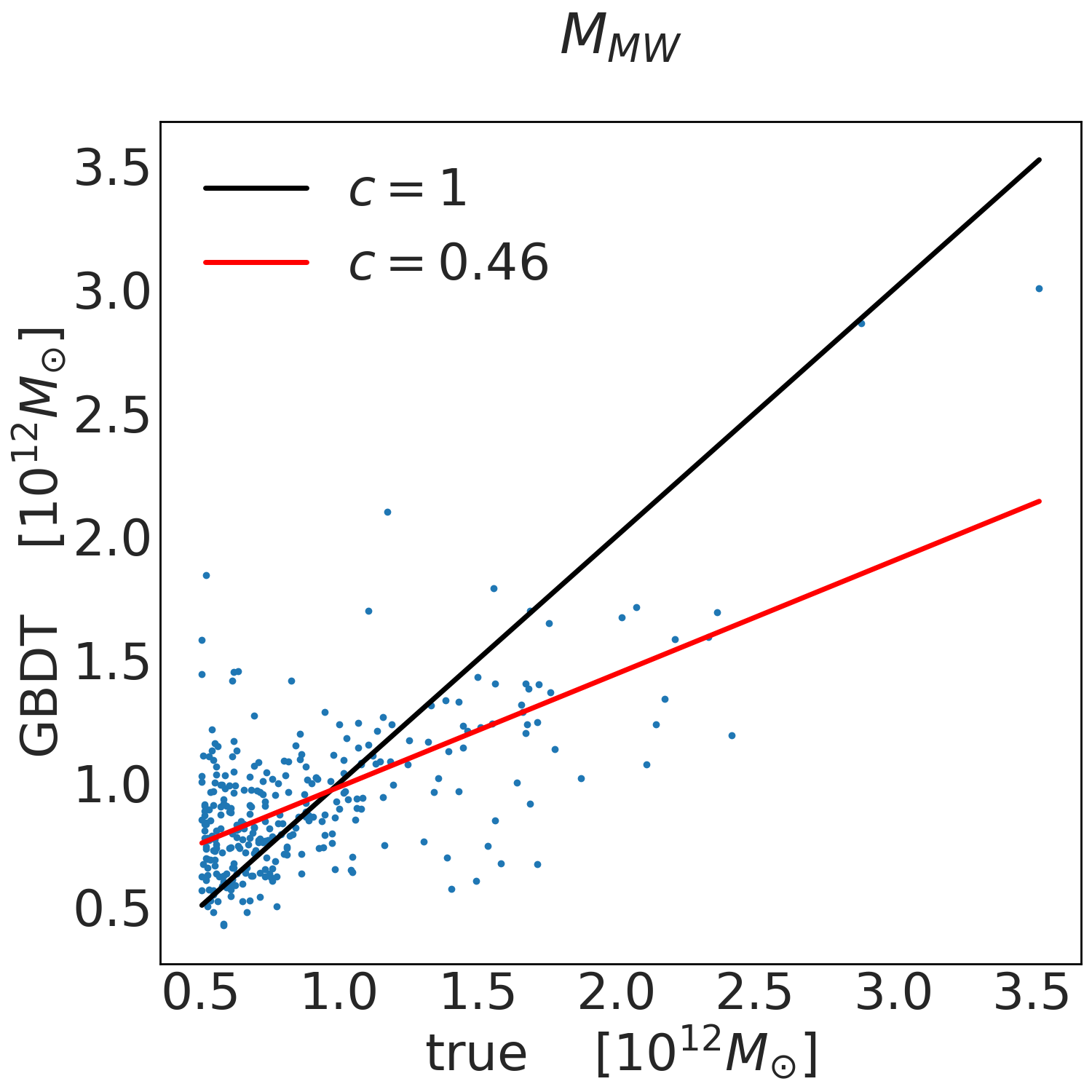}  
		\includegraphics[width=0.4\textwidth]{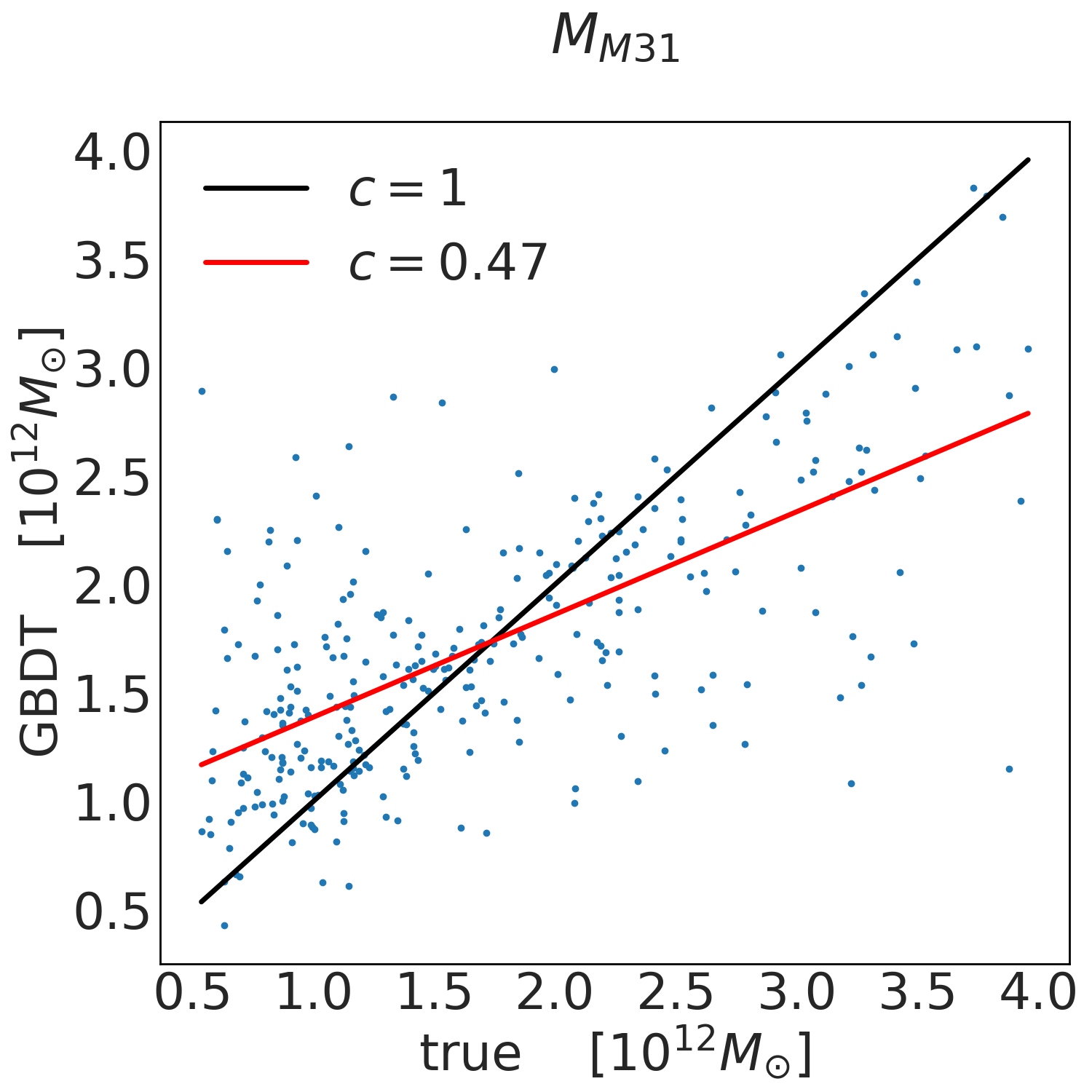}  
		\includegraphics[width=0.4\textwidth]{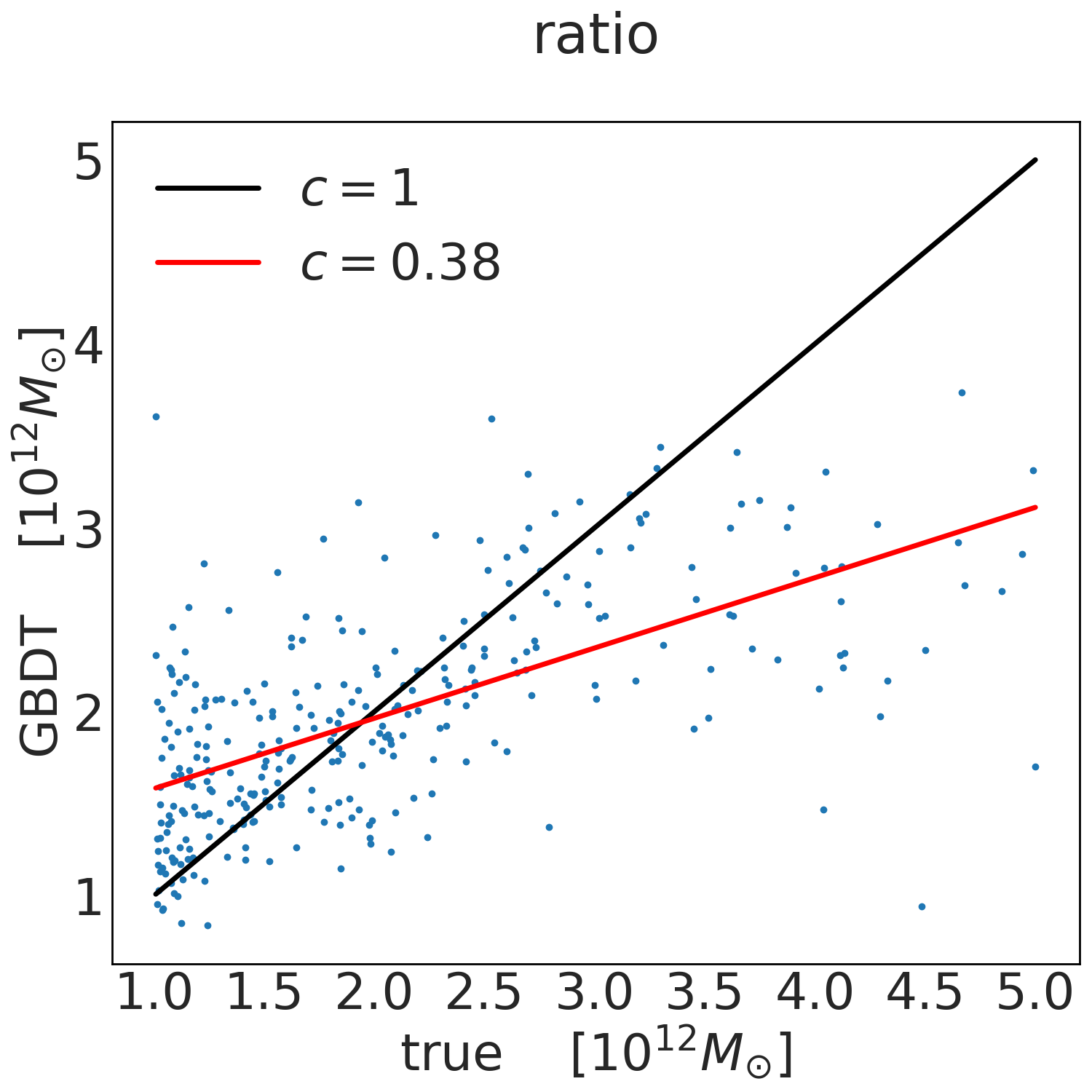}  
	\end{center}
\caption{
The GBDT algorithm applied to the mock LGs of the validation set: The \mlg\ (upper-left), \mmw\ (upper-right), \mm31\ (lower-left) and the ratio \mmto\ / \mmw\ (lower-right) panels show scatter-plots of the predicted vs. the true values.
		The  red solid line show the best fit slope ($c$) obtained here for the predicted vs the true values. The case of slope of unity is shown for reference (blue solid line).
}
\label{fig:scatter-plots}
\end{figure*}

\begin{figure*}
\begin{center}
		\includegraphics[width=0.4\textwidth]{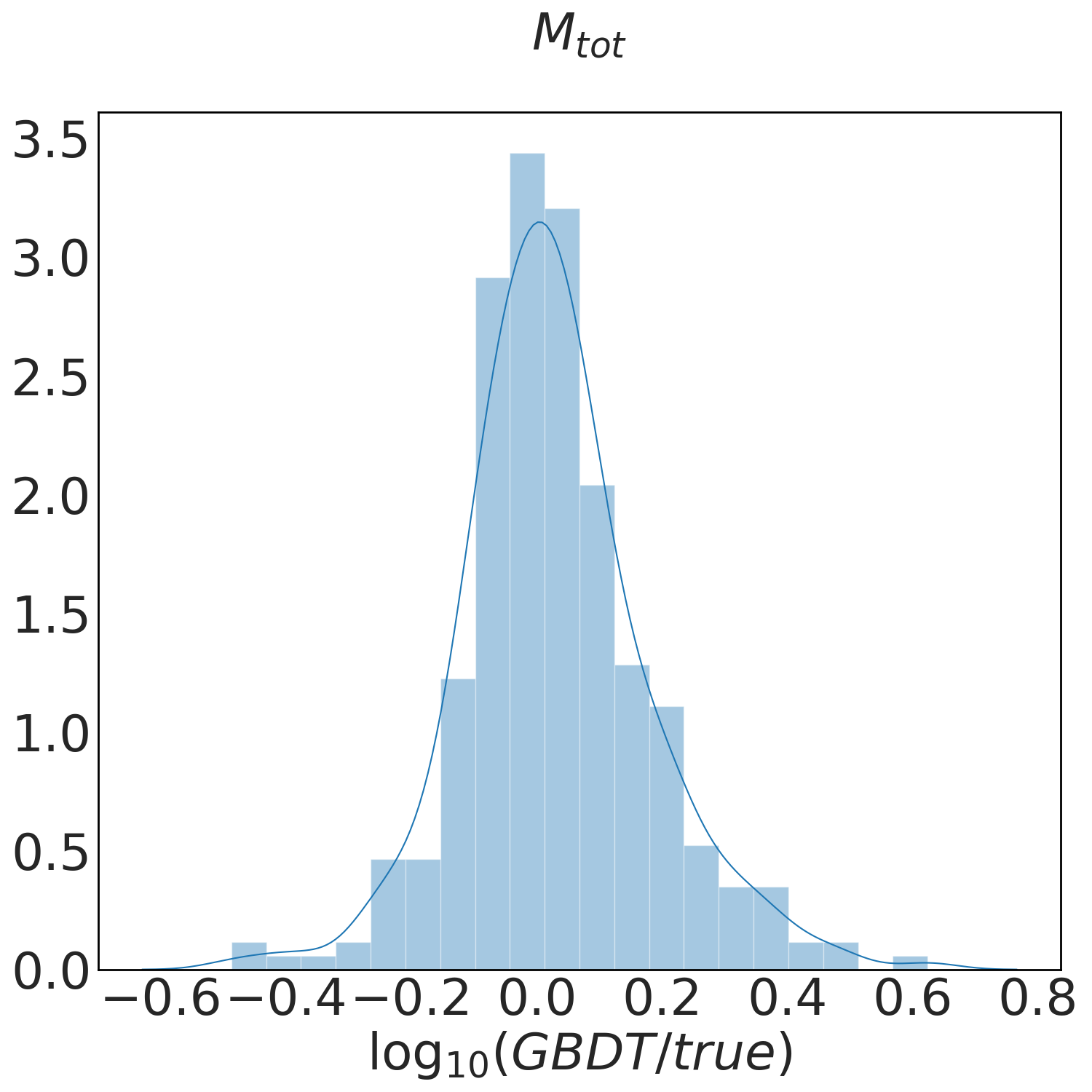}  
		\includegraphics[width=0.4\textwidth]{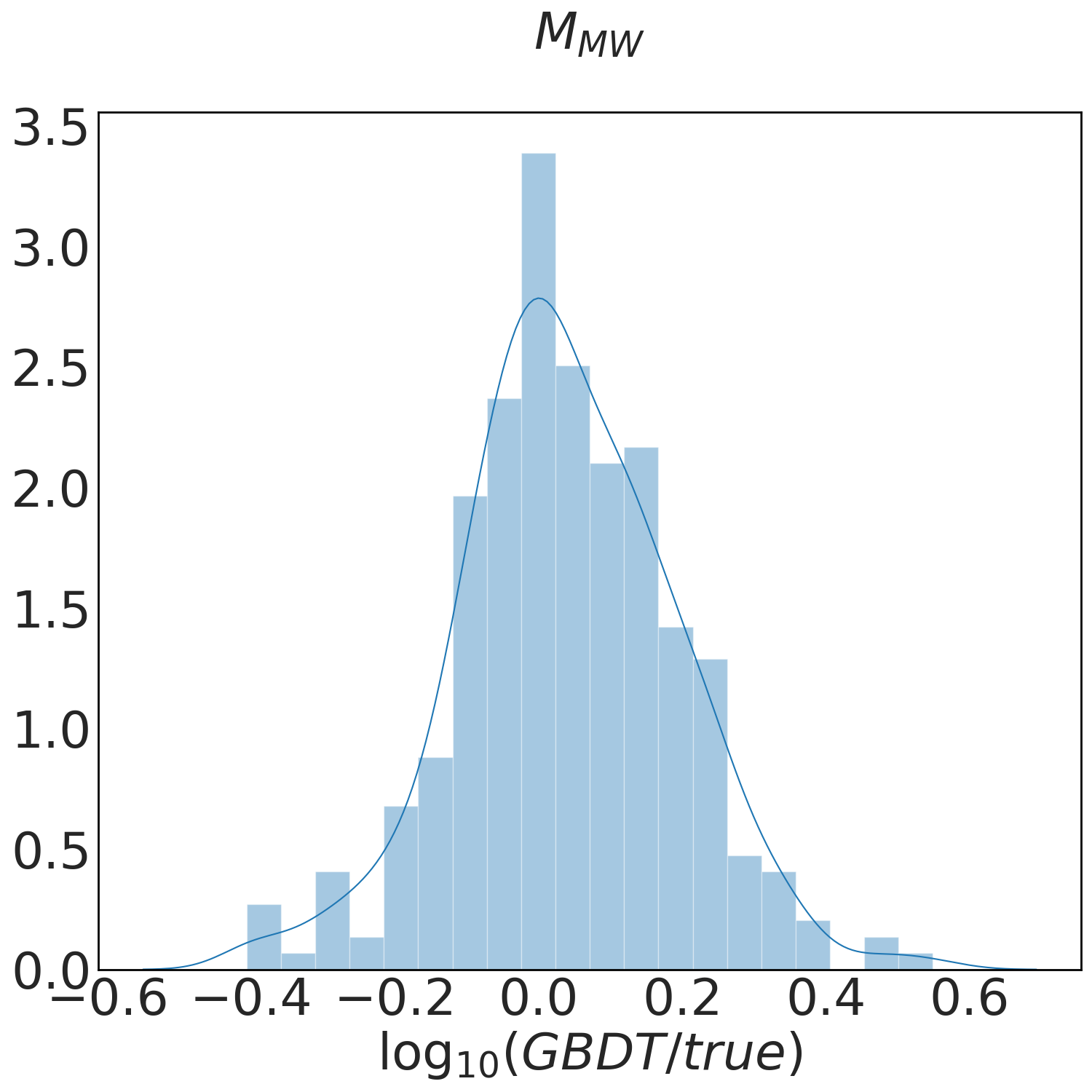}  
		\includegraphics[width=0.4\textwidth]{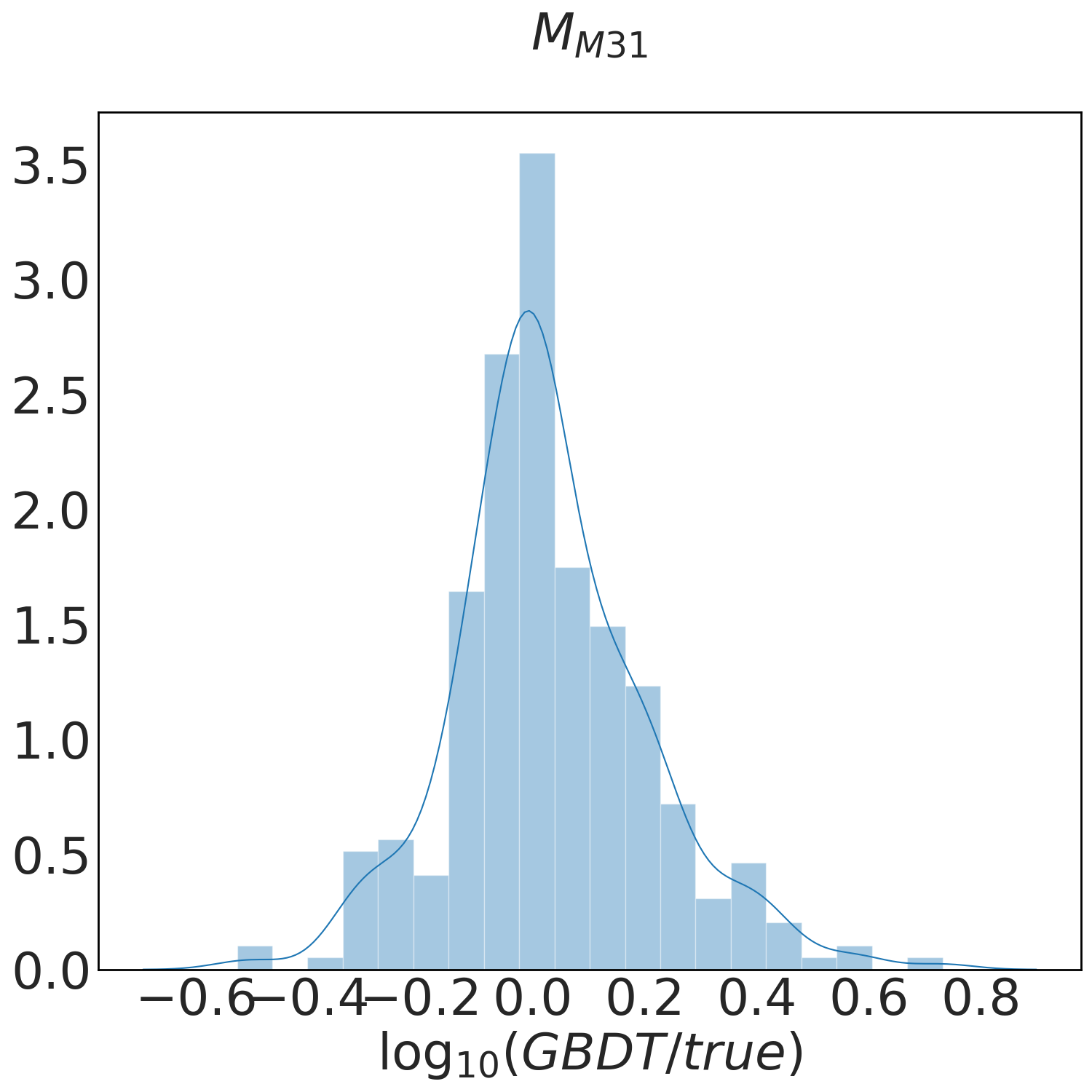}  
		\includegraphics[width=0.4\textwidth]{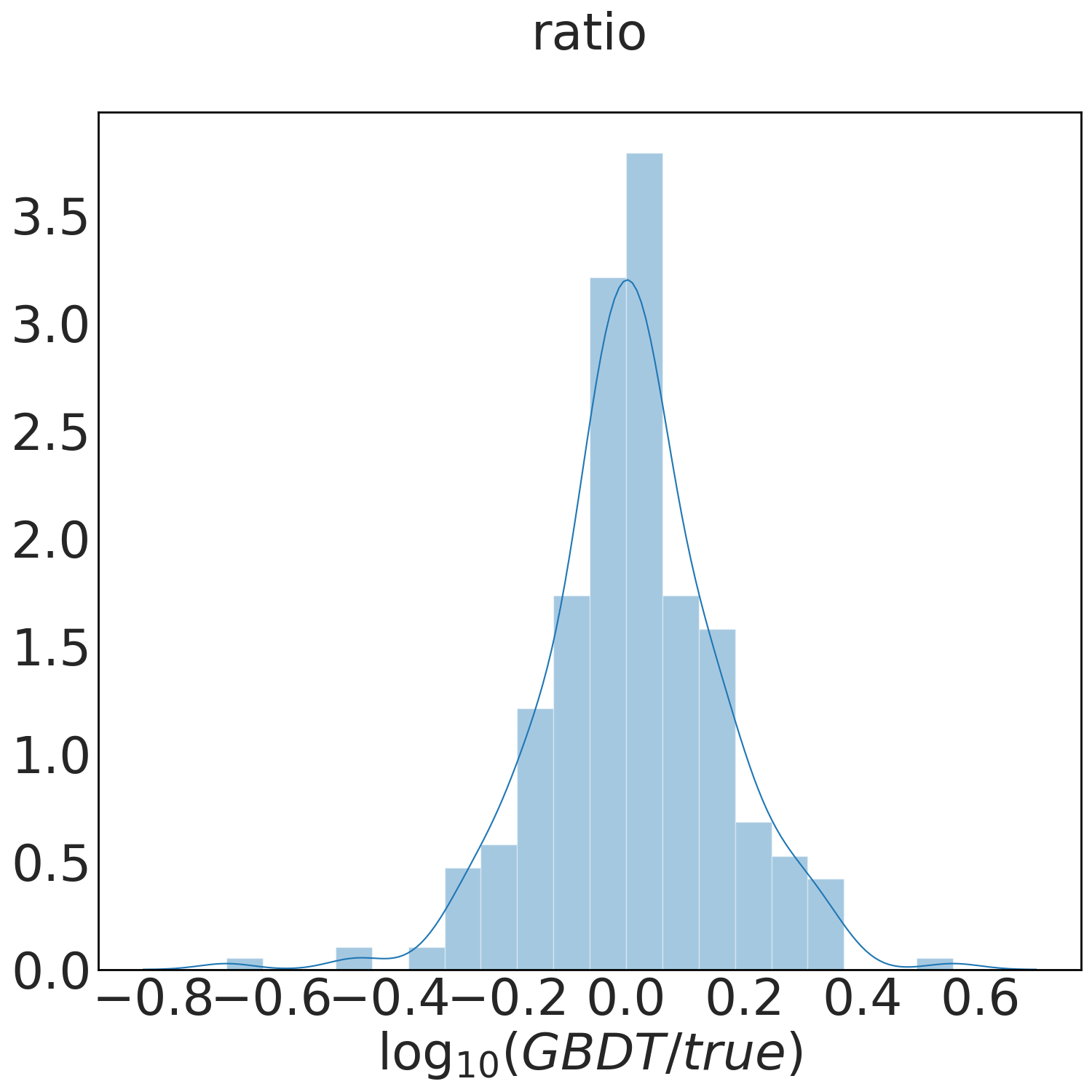} 
\end{center}
\caption{The GBDT algorithm applied to the mock LGs of the validation set: The \mlg\ (upper-left), \mmw\ (upper-right), \mm31\ (lower-left) and the ratio \mmto\ / \mmw\ (lower-right) panels show the histogram of the $\log_{10}$ of the ratios of the predicted over  the true masses and of the predicted over the true values of $r_M$.
The solid lines on the histograms are the smoothed kernel density estimates of the distributions.
}
\label{fig:histograms}
\end{figure*}

\begin{table}
\begin{center}
	\caption{Slope $c$ of the predicted versus true value; median $\mu$ and scatter $\sigma $
	for the $(\hat{y} / y )$ computed for \mlg , \mmw, \mmto\ and $r_{M}$. The errors have 
	been computed by a bootstrapping method to test for the robustness of the results. 
	The deleted text has been moved to the main body of the text.}
\label{tab:ml}
\begin{tabular}{ccccc}
\hline
	$\quad$  & \mlg    & \mmw & \mmto  & $r_{M}$\\
\hline
	$c$      & $0.47\pm0.03$  & $0.46\pm0.03$  & $0.47\pm0.03$  & $0.38\pm0.04$ \\
	$\mu$    & $0.006\pm0.002$ & $-0.12\pm0.01$ & $-0.07\pm0.01$ & $0.011\pm0.005$ \\
	$\sigma$ & $0.23\pm0.02$  & $0.46\pm0.07$  & $0.39\pm0.03$  & $0.15\pm0.03$ \\
\hline
\end{tabular}
\end{center}
\end{table}

The sample of 2148 mock LGs are split into 1611 training and 537 validation sets.  
The GBDT hyperparameters are gauged and optimized by monitoring over the mock LGs of the training set of 
the slope ($c$) between predicted versus true values, $y_{pred} = c y_{true}$ and the 
the median ($\mu$) and the standard deviation  ($\sigma$)  of the residual $y_{pred} - c y_{true}$. Here $y$ stands of the $\log_{10}$ of $M_{MW}$, $M_{M31}$ and $M_{LG}$ and of the mass ratio $r_M$. The  resulting estimation is then then tested against the validation set.

The uncertainties in the GBDT estimated slope, median and standard deviation are obtained according to a boot strapping technique. Accordingly, 200 pairs are randomly drawn from the 537 validation data. For each draw, a value for the slope $c$, the   median $\mu$   and standard deviation ($\sigma$) are computed.  The values reported in Table \ref{tab:ml} are the means and the standard deviations computed over these 200 subsabmples.

Our main results are obtained for $n_{tree}=250$, $n_{depth}=10$ and $n_{leaf}=5$.
These are presented in  Table \ref{tab:ml}
and in Figs. \ref{fig:scatter-plots} and \ref{fig:histograms}.
Summary of these results follow: 
The correlation between the true and predicted value $c$ is in the range  $0.37-0.49$, in the case of \mlg\ this value is comparable with $c\approx 0.6$ quoted by
\citet{McLeod:2017} as obtained with an artificial neural network algorithm (ANN).
The medians  ($\mu$) of the ratio of the log of the predicted to the true value,of  \mlg\ %\mmto\  
and the mass ratio is essentially unbiased with a relatively small scatter. The predictions of the masses of the individual MW and M31 galaxies are  less robust, yet the medians of the distribution of the log of the masses are well within the scatter. The mass of M31 is somewhat better predicted than that of the MW galaxy. 
The nature of the LG model  assumed here readily explains these shortcomings. The masses of the individual galaxies are associated here those of the corresponding halos, identified here by AHF halo finder - a definition that relies on the virial theorem. The mass function of such halos in the mass range defined by the LG model is a strongly decreasing function of the mass. Given the fixed range of masses associated with the MW and M31 galaxies, one expects the GBDT to somewhat underestimate the masses of the individual galaxies, and that of the MW, which by definition is the less massive one, to be more affected by that bias. It can be removed by properly revising the LG model, but this lies beyond the scope of the paper.

\section{Application to the actual Local Group}
\label{sec:actual_LG}

\subsection{Monte Carlo sampling of errors}
\label{subsec:MC}

\begin{table}
\begin{center}
	\caption{Mean and standard deviation values for $r$, \vrad, and \vtan  used to generate the MC data sample.}
\label{tab:intervals}
\begin{tabular}{cccc}
\hline
	$\quad$ & $\mu$ & $\sigma$ & unit \\
\hline
	$r$ & 770 & 40 & \kpc\\
	\vtan & 57 & 35 & \kms \\
	\vrad & 105 & 5 & \kms \\
\hline
\end{tabular}
\end{center}
\end{table}
 
The input parameters of the GBDT algorithm are the following observables: the distance ($r$), relative radial and  (\vrad) and tangential (\vtan) velocities of the MW and M31 galaxies. Observational estimates of these observables and their observational  uncertainties are presented in Table \ref{tab:intervals}. The estimations of the distances and radial velocity are taken from \citet{Marel:2012} and of the tangential velocity  from \citet{Marel:2019} (using \emph{Gaia} data).

The uncertainties in the inferred GBDT  masses and their ratio are estimated by means of the forward modeling. Namely,  
given the data of Table \ref{tab:intervals}  an ensemble of $10^4$ Monte Carlo realizations of the possible values of input parameters of the actual LG has been generated. The realizations are generated by assuming normal distribution of the uncertainties, and are done under the assumption of uncorrelated errors.  The GBDT is then applied to that ensemble of possible values of input parameters. The key assumption that is made here is that the uncertainties quoted in Table \ref{tab:intervals} are statistically uncorrelated.

\subsection{Results}
\label{subsec:LG_results}

\begin{figure*}
	\begin{center}	
		\includegraphics[width=0.4\textwidth]{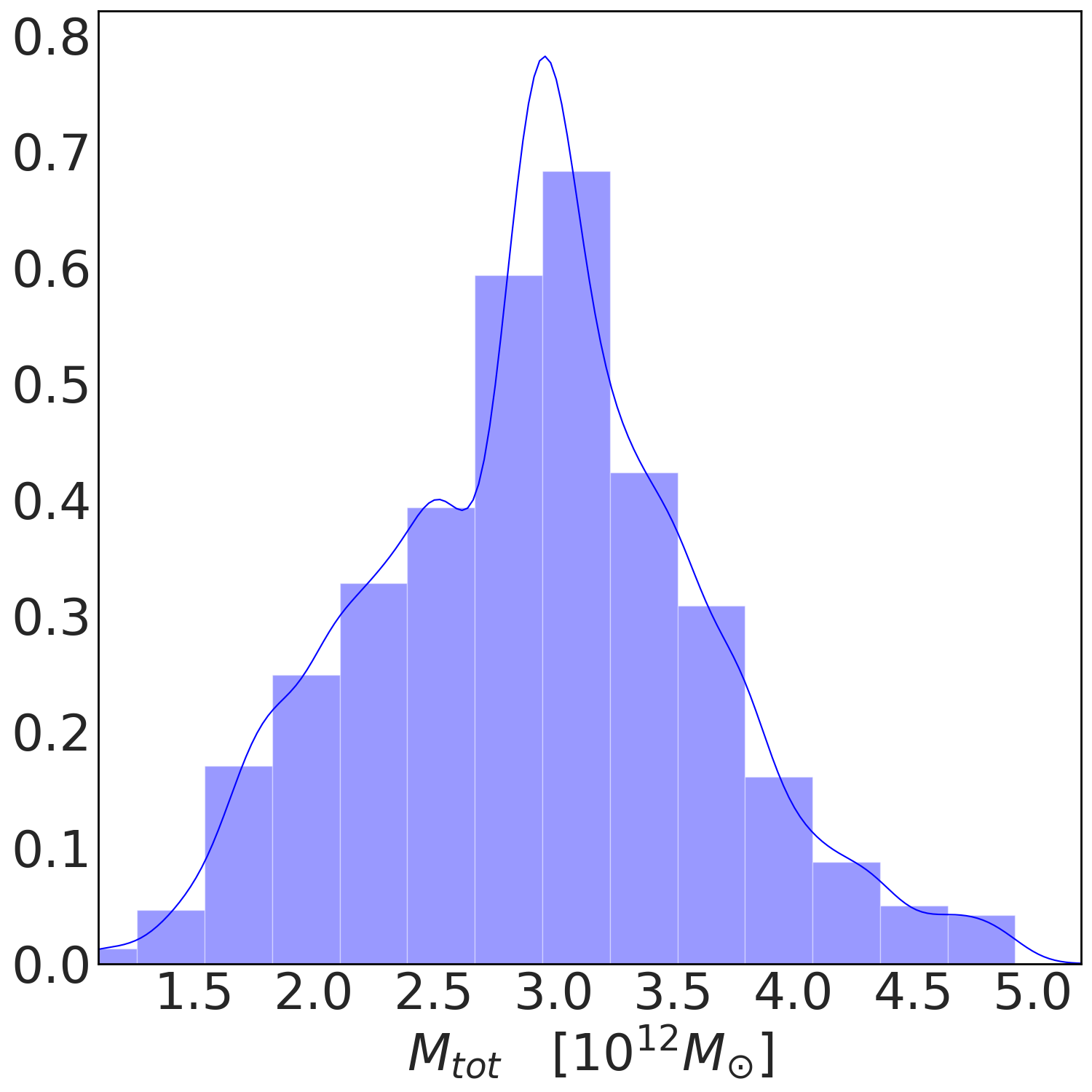} 
		\includegraphics[width=0.4\textwidth]{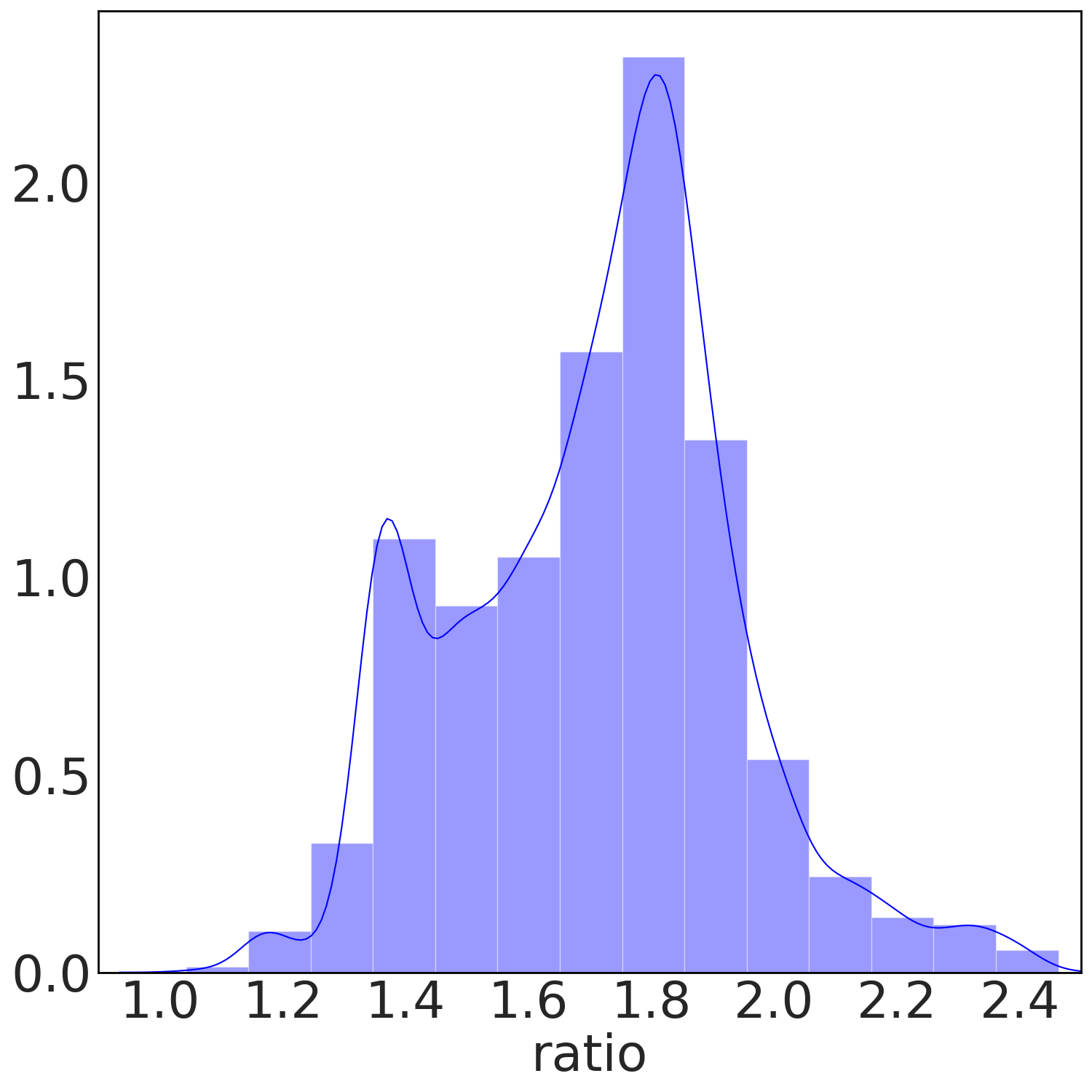}	
	\end{center}
		\caption{GBDT estimate of the total mass (left panel; note the linear scale of the mass) and of the ratio of $M_{M31} / M_{MW}$ (right panel). The histogram correspond to the distribution of the Monte Carlo realizations of the observational uncertainties. Smoothed kernel density estimates are plotted on top with solid lines.}
\label{fig:mc_LG_rM}
\end{figure*}

\begin{figure}	
	\begin{center}	
	\includegraphics[width=0.45\textwidth]{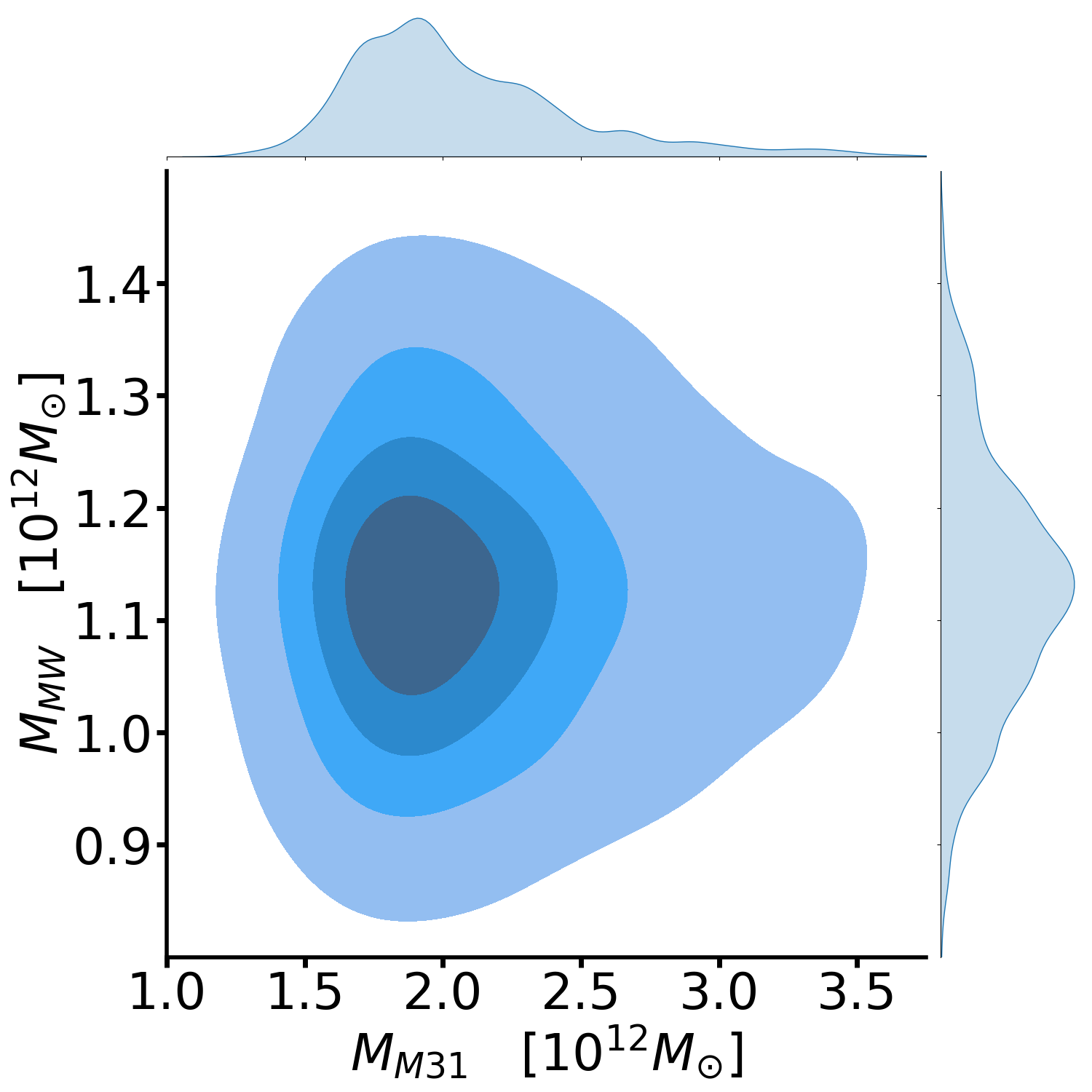}
	\end{center}
\caption{Joint smoothed kernel density estimate for the GBDT estimate of MW and M31 masses obtained with
 mock Monte Carlo realizations of the observational uncertainties. The univariate smoothed estimates are also projected.}
\label{fig:mc_MW_M31}
\end{figure}

The ensemble of $10^4$ Monte Carlo realizations of the possible values of the  LG  (\mlg), the Milky Way   (\mmw), and the M31  (\mmto)  and the   ratio of the masses of the two galaxies $r_{M}$ is used to calculate the posterior distribution of these masses and ratio of the masses.   The posterior distribution functions - given the observables, the LG model and the \LCDM\ model - are presented in Figs. \ref{fig:mc_LG_rM} and \ref{fig:mc_MW_M31}. The medians and the lower and upper quartiles are  given in Table \ref{tab:mc}. The main results of the GBDT analysis are presented here. 
No attempt is made here to convert all the masses quoted in Table \ref{tab:mc} to a unified definition. The halo mass, taken from the analysis of  numerical simulations, used here and also by \citet[][]{McLeod:2017}  and  \citet[][]{Lemos:2021} 
is $M_{200}$ as defined by the AHF halo finder.

\noindent
\textbf{Feature importance:} 
A unique feature of the the GBDT algorithm is its ability to rank  features, i.e. the input parameters,  by their importance. Namely  the weight each feature  carries 
in producing the final output value is estimated. Our first result is  that the three features have the same weight $\approx 0.33$ (Table \ref{tab:fi}), and therefore are equally important for the determination  of the various masses and the mass ratio reported here.
This is consistent with the results of \citet{Carlesi:2017a}, which explored the impact of the different LG models on mass predictions, finding that
modification in the priors for \vtan, \vrad\ and $r$ lead to comparable changes in the results.

\begin{table}
\begin{center}
	\caption{ \emph{Feature importance} for the four GBDT models.}
\label{tab:fi}
\begin{tabular}{cccc}
\hline
	$\quad$ &  $r$ & \vrad & \vtan \\
\hline
	\mlg        & 0.33 & 0.35 & 0.32\\
	\mmw       & 0.35 & 0.33 & 0.32\\
	\mmto       & 0.34 & 0.35 & 0.31 \\
	$r_{M}$    & 0.33 & 0.34 & 0.34 \\
\hline
\end{tabular}
\end{center}
\end{table}

\begin{table}
\begin{center}
	\caption{Comparison between the GBDT results and some of the other methods found in the literature to estimate \mlg, \mmw, \mmto\ and the mass ratio of M31 to MW.
	We can see that the GBDT method provides estimates that compare favourably with previous results and make it suitable for a wide range of different applications.}
\label{tab:mc}
 \begin{tabular}{cc}
\hline
	Reference & \mlg $[10^{12}\Msun]$\\
\hline
	Present paper     & $3.31^{+0.53}_{-0.45}$ \\ 
	\citet[][]{Marel:2012} & $3.17\pm0.57$  \\
	\citet[][]{Gonzalez:2014} & $4.17^{+1.45}_{-0.93}$ \\
	\citet[][]{Penarrubia:2016} & $2.64^{+0.42}_{-0.38}$ \\
	\citet[][]{Carlesi:2016b} & $3.47^{+1.16}_{-0.83}$\\
	\citet[][]{McLeod:2017} & $3.6^{+1.3}_{-1.1}$ \\
	\citet[][]{Lemos:2021} & $4.6^{+2.3}_{-1.8}$ \\
\hline	
\\
\hline
	Reference & \mmw $[10^{12}\Msun]$\\
\hline
	Present paper     & $1.15^{+0.25}_{-0.22}$ \\ 
	\citet[][]{Battaglia:2005} & $1.2\pm0.5$  \\
	\citet[][]{McMillian:2011} & $1.26\pm0.24$ \\
	\citet[][]{Boylan-Kolchin:2013} & $1.6^{+0.8}_{-0.6}$ \\
		\citet[][]{Gonzalez:2014} & $1.14^{+1.19}_{-0.41}$ \\
	\citet[][]{Diaz:2014} & $0.8\pm0.5$ \\
	\citet[][]{McMillian:2016} & $1.3\pm0.3$ \\
	\citet[][]{Carlesi:2016b} & $0.82^{+0.33}_{-0.38}$\\
	\citet[][]{Fragione:2016} & $1.2 - 1.7$ \\
	\citet[][]{Watkins:2019} & $1.28^{+0.97}_{-0.48}$ \\
\hline	
\\
\hline
	Reference & \mmto $[10^{12}\Msun]$\\
\hline
	Present paper     & $2.01^{+0.65}_{-0.39}$ \\ 
	\citet[][]{Fardal:2013} & $1.99^{+0.51}_{-0.41}$ \\
	\citet[][]{Diaz:2014} & $1.7\pm0.3$ \\
	\citet[][]{Carlesi:2016b} & $2.48^{+0.80}_{-0.56}$\\
	\citet[][]{Penarrubia:2016} & $1.33^{+0.97}_{-0.48}$ \\
\hline	
\\
\hline
	Reference & \mmto / \mmw \\
\hline
	Present paper     & $1.75^{+0.54}_{-0.28}$ \\ 
	\citet[][]{Baiesi:2009} & $2.5\pm0.5$  \\
	\citet[][]{Diaz:2014} & $2.3^{+2.1}_{-1.1}$ \\
\hline	

\end{tabular}
 \end{center}
\end{table}

\noindent
\\
\\
\textbf{Total mass:}
The GBDT estimate  of \mlg $= 3.31 ^{+0.79}_{-0.67}$ (all masses are in  $10^{12}$\Msun\ units) is in  a very good agreement with the results of \citet{Marel:2012} and 
\citet{McLeod:2017}, while it
lies between the smaller results of \citet{Penarrubia:2016} and \citet{Diaz:2014}  and the higher ones by \citet{Gonzalez:2014, Fattahi:2016a} and \citet{Lemos:2021} (see 
the first block of \Tab{tab:mc})
\\

\noindent
\textbf{Milky way mass:} The GBDT estimation  of \mmw$=1.15{^{+0.25}_{-0.22}} $ is  in a very good agreement with the \emph{Gaia} value of \citep{Watkins:2019} 
as well as earlier estimates by \citet{Battaglia:2005, McMillian:2011, Gonzalez:2014, McMillian:2016} within the same mass range.
The present estimate  is halfway between the results of e.g. \citet{Gibbons:2014, Carlesi:2016b}  
  and the ones above predicted from the analysis of high velocity stars of \citep[]{Fragione:2016} or the value of implied by a bound orbit for Leo I \citep{Boylan-Kolchin:2013}, 
  reported in the second block of \Tab{tab:mc}.
\\

\noindent
\textbf{M31 mass:}
The GBDT estimation  of \mmto$=2.01^{+0.65}_{-0.39}$ is in good agreement with   the results obtained by \citet{Fardal:2013}, and \citet{Diaz:2014} as well as the results of \citet{Carlesi:2017a} obtained for a non radial motion (i.e. high \vtan ) of M31 with respect to MW (see \Tab{tab:mc}).
\\

\noindent
\textbf{Mass ratio:}
The GBDT model estimate of  $r_M=1.75^{+0.54}_{-0.28}$   is  in agreement with that of  \citet{Baiesi:2009},
estimated studying the tidal perturbations of M31 to the LG and \citet{Diaz:2014} obtained by balancing the angular momentum at the LG center of mass, as can be seen 
in the last block of \Tab{tab:mc}.
\\

\section{Discussion}
\label{sec:disc}

A supervised machine learning algorithm, the gradient boosted decision trees, is  derived so as to estimate   the masses of the Local Group and its main components, the Milky Way and M31 galaxies, and their mass ratio, from the observed distance and relative velocity of the two. The algorithm is trained and validated against a sample of 2148  mock LGs drawn from   a set of five DM only $N$-body simulations  \citep[see][]{Carlesi:2019}
with $1024^3$ particles within a 100 \hMpc\ box with Planck-I cosmological parameters \citep{Planck:2013}.
The halo catalogs were obtained with  AHF  halo finder \citep{Knollmann:2009}. The selection of the mock LGs is based on a LG model introduced here. 

A summary of the main results is: a. Total mass of the LG,  \mlg $= 3.31 ^{+0.79}_{-0.67}$ (median and 1st and 3rd quartiles; masses are in  $10^{12}$\Msun\ units); b.  Mass of the MW galaxy, \mmw$=1.15{^{+0.25}_{-0.22}} $;  c. Mass of the M31 galaxy \mmto$=2.01^{+0.65}_{-0.39}$; d. The mass ratio  $M_{M31} / M_{MW} =1.75^{+0.54}_{-0.28}$; e. Features importance of the three input parameters, distance, relative radial and tangential velocities, is roughly equal. The estimated masses of the two individual galaxies and of the total mass of the LG are in reasonable to very good agreement with other published estimates and as such we do not expect them to be controversial. Our estimate of the the mass ratio stands out to some degree. The issue of possible approximate equality of the masses of the Milky Way and M31 galaxies is open and somewhat disputed. Studies based on the timing argument for example compute a total LG mass, while studies based on the kinematics of MW stars infer a halo mass via extrapolation to the virial radius. Taken at face value, independent individual measurements (or estimations) of the MW and M31 halo masses are unable to either constrain the mass ratio, nor make firm statements on which is the more massive halo. With our approach we cannot argue which galaxy is more massive but predict that their mass ratio is  $1.75^{+0.54}_{-0.28}$.

Traditionally the issue of the mass of the LG as a whole or of masses of the main objects that constitute the LG, has been dealt within a dynamical framework. Much of that work has been inspired by the  \emph{timing argument} model, which simplifies the complex dynamics and mass assembly history of the LG as the classical two-body problem, thereby  envisaging the MW and M31 galaxies as two point-like particles  conserving their  orbital energy and angular momentum and their  individual initial masses. The `two-body  problem' model does not capture the  complexity of the dynamics of the LG (as mentioned in the Introduction). This has  led to a  new approach, of estimating the mass of the LG by means of ML techniques applied to simulated LG-like objects 
\citep{McLeod:2017,2020JCAP...09..056M,Lemos:2021}. These studies gave up the hope on amending the \emph{timing argument} model by a rigorous dynamical treatment and opted for ML techniques. Yet, the input parameters extracted from the simulations have been inspired by the simple analytical model - mostly the distance and relative velocities of the MW and M31-like objects. The success of those ML studies in estimating the  mass of the LG is in some sense not surprising, as they operate withing the same framework as the \emph{timing argument} model.  The success  of our current analysis in estimating the individual masses of the MW and M31 galaxies is surprising, as nothing in the dynamics of the two-body problem and in the definition of the LG Model hints towards a sensitivity to these individual masses. Recent developments aiming at gaining insight to the interpretability and explainability of the results of ML methods might shed more light on the problem and add to our understanding  of the dynamics of the LG  
\cite[e.g.][]{e23010018,2020arXiv201201805M}. This unfortunately  lies outside the scope of our paper.

We wish to conclude with a general assessment of the application of supervised ML in general and the GBDT algorithm in particular to cosmological problems and to the estimation of the parameters of given cosmological objects like the LG. The analysis presented here has been formulated within a clear framework: a. The standard  cosmological   model (\LCDM) is assumed; b. A LG model, which defines what   a mock LG is; c. Observational uncertainties of the input parameters are assumed to be normally distributed. The GBDT algorithm is constructed as a multi-parameter non-linear fitting procedure, whose parameters are fixed by following the prior assumptions (\LCDM\ and the LG model). The  algorithm is then applied to the ensemble of the Monte Carlo realization of the input observational data and its uncertainties, thereby it constitutes a non-linear mapping from the prior probability distribution function of the input data to the posterior distribution functions of the predicted parameters. It follows that the GBDT algorithm in particular and supervised ML tools 
constitute  a framework that is similar in its nature to the Bayesian approach, as all the conclusions and results presented here are strictly valid only under the assumptions of the \LCDM\ and the LG models and  given the observed values of the distance and relative velocity of the MW and M31 galaxies.

\section*{Addendum}
After the submission of the paper we came across the work of  \citet{2021arXiv211114874V} who have estimated the  masses of the MW and M31 galaxies by means of an ML technique, the graph neural networks, from a set of hydrodynamic cosmological simulations. In that study the issue of whether the MW and M31 galaxies are members of the LG is not considered. The resulting estimated mass of the M31  of that study is in very good agreement with ours. As for the MW galaxy, \citet{2021arXiv211114874V} have two different ways of estimating the mass, with and without including the velocities of satellites of the studied galaxies in their input data. The MW mass of the 'without the velocities' case is  in a very good agreement  with   our estimation while in the other case the estimations are more than  1 sigma apart. In all cases the error bars  presented here are  smaller by roughly a factor of 2 or larger than those of \citet{2021arXiv211114874V}.

\section*{Acknowledgements}
This work has been done within the framework of the Constrained Local UniversE Simulations (CLUES) simulations. 
YH has been partially supported by the Israel Science Foundation grant ISF 1358/18.
NIL   acknowledges financial support of the Project IDEXLYON at the University of Lyon under the Investments for the Future Program (ANR-16-IDEX-0005). 

\section*{Data availability}
The code used for the GBDT, the analysis and the plots is freely available on the lead author's GitHub repository: https://github.com/EdoardoCarlesi/MLEKO.
The simulation data used for the training of the GBDT algorithm can be obtained by contacting the lead author at ecarlesi83@gmail.com.

%%%%%%%%%%%%%%%%%%%%%%%%%%%%%%%%%%%%%%%%%%%%%%%%%%%

\bibliographystyle{mn2e}
\bibliography{biblio}

\bsp

\label{lastpage}

\end{document}